\documentclass[aps,showpacs,preprintnumbers]{revtex4}
\usepackage{amsmath}
\usepackage{bm}

\input{tcilatex}

\begin{document}

\title{Gauge transformations are not canonical transformations }
\author{A.T.Suzuki}
\affiliation{Instituto de F\'{\i}sica Te\'orica-UNESP,\\
Rua Pamplona 145 -- 01405-900 S\~ao Paulo, SP -- Brazil}
\author{J.H.O.Sales}
\altaffiliation[ ]{Funda\c{c}\~ao de Ensino e Pesquisa de Itajub\'a \\
Av. Dr. Antonio Braga Filho, CEP 37501-002, Itajub\'a, MG.}
\affiliation{Instituto de Ci\^{e}ncias, Universidade Federal de Itajub\'{a}, CEP
37500-000, Itajub\'{a}, MG, Brazil. }
\date{\today}

\begin{abstract}
In classical mechanics, we can describe the dynamics of a given system using either the Lagrangian formalism or the Hamiltonian formalism, the choice of either one being determined by whether one wants to deal with a second degree differential equation or a pair of first degree ones. For the former approach, we know that the Euler-Lagrange equation of motion remains invariant under additive total derivative with respect to time of any function of coordinates and time in the Lagrangian function, whereas the latter one is invariant under canonical transformations. In this short paper we address the question whether the transformation that leaves the Euler-Lagrange equation of motion invariant is also a canonical transformation and show that it is not.
\end{abstract}

\maketitle







\section{Introduction}

In considering the subject of gauge transformations and canonical transformations, our attention was drawn to the work of T.L.Chow \cite{Chow} where he claims that a gauge transformation that leaves the Euler-Lagrange equations of motion invariant is also a canonical transformation that leaves the Hamilton's equations of motion invariant. In the aftermath of this review, our conclusion is radically different from his, and demonstrate that the gauge transformation that leaves the Euler-Lagrange equations of motion invariant {\bf is not} a canonical transformation.
   
\section{Gauge invariance in the Lagrangian formalism}

For a dynamical system, there exists a well-known property which states that given two Lagrange functions $L(q,\dot q,t)$ and $L'(q,\dot q,t)$ that differs only by a total derivative with respect to time of any function of coordinates and time, that is,
\begin{equation}
L'(q, \dot q, t) = L(q, \dot q, t) + \frac{d}{dt}f(q,t), \label{gauge_transf}
\end{equation}
where as usual we denote the generalized coordinates as $q\equiv \{q_i\} = \{q_1,\,q_2,\,...q_N\}, \:\:\:i=1,\,2,\,...N$ and $\dot q \equiv \displaystyle\frac{dq}{dt}$. Since the presence of total derivative with respect to time that appears on the right-hand side of (\ref{gauge_transf}) does not affect the Euler-Lagrange equations of motion, one can loosely call it a ''gauge'' transformation, in the sense that equations of motion are invariant under such a transformation.

On the other hand, equations of motion are also invariant under the so-called {\em point transformations} in which the choice for the generalized coordinates $q$ that defines uniquely the position of a given system in space is modified under transformations of the type
\begin{equation}
Q=Q(q,t). \label{point_transf}
\end{equation}

This transformation leaves invariant not only the Euler-Lagrange equations of motion but the Hamilton (canonical) equations of motion as well \cite{Landau}. So, {\em point transformations} are canonical transformations because they preserve the canonical equations of motion. 

A more general type of transformations is possible, since in the Hamiltonian formalism the momenta $p\equiv \{p_i\}$ are independent variables, so that an ampler class of transformations looks like 
\begin{eqnarray}
Q&=&Q(p,q,t), \\
P&=&P(p,q,t). \label{ampler_transf}
\end{eqnarray}

Not all transformations of the type (\ref{ampler_transf}) preserve the canonical form of Hamilton's equations of motion. For those transformations which do preserve the canonical form, they receive the name of {\em canonical transformations}.

However, the question that interest us now is whether the gauge transformation (\ref{gauge_transf}) is a canonical transformation. The answer is no, as we shall shortly see.

First of all let us check the invariance of Euler-Lagrange equations of motion under (\ref{gauge_transf}). The action is
\begin{eqnarray}
S'& = & \int_{t_i}^{t_f} L'(q,\,\dot q,\,t)\,dt \nonumber \\
  & = & \int_{t_i}^{t_f} L(q,\,\dot q,\,t)\,dt + \int_{t_i}^{t_f} \frac {df(q,t)}{dt}\,dt \nonumber \\
  & = & S + f(q_f,\,t_f) - f(q_i,\,t_i)
\end{eqnarray}
where $q_f$ and $q_i$ are the coordinates for times $t_f$ and $t_i$ respectively. Euler-Lagrange equations of motion are obtained by varying this action, and from $\delta S'=0$ and $\delta S=0$ it follows immediately that Euler-Lagrange equations of motion remain invariant under (\ref{gauge_transf}), since the extremes $q_f$ and $q_i$ are fixed points. 

\section{Canonical transformations}

In the previous section we have stated that not all transformations of the type (\ref{ampler_transf}) are canonical transformations. Here we shall determine which conditions or constraints we must have in order to have them as canonical transformations. For this purpose, we start off by considering that Hamilton's equations of motion can be obtained from the principle of least action in the form
\begin{equation}
\delta \: \int \left(\sum_i p_i\,dq_i - H\,dt\right) = 0 \label{Hamilton}
\end{equation}
where all coordinates and momenta are varied {\em independently}. In order for the new variables $P$ and $Q$ to satisfy the Hamilton's equation of motion, the principle of least action must be also valid here, i.e.,
\begin{equation}
\delta \: \int \left(\sum_i P_i\,dQ_i - H'\,dt\right) = 0 \label{Hamilton1}
\end{equation}

The two principles (\ref{Hamilton}) and (\ref{Hamilton1}) will be equivalent if and only if the two integrands differ by a total derivative of an arbitrary function $G=G(p,q,t)$ so that the difference between the two integrals will be a difference between the function $G$ in the upper and lower limits of integration. Then their independent variation with respect to coordinates and momenta vanishes and we have
\begin{equation}
\sum_i p_i\,dq_i - H\,dt = \sum_i P_i\,dQ_i-H'\,dt+dG.
\end{equation}

From this, we can write
\begin{equation}
dG = \sum_i p_i\,dq_i-\sum_i P_i\,dQ_i+(H'-H)\,dt,
\end{equation}
so that
\begin{eqnarray}
p_i & = & \frac{\partial G}{\partial q_i} \\
P_i & = & \frac{\partial G}{\partial Q_i} \\
H'= & = & H+\frac{\partial G}{\partial t}
\end{eqnarray}

Therefore, here $G=G(q,Q,t)$ is the generating function of the canonical transformation. We can proceed by using Legendre transformations to work out other canonical transformations from generating functions such as $G=G(q,P,t)$, $G=G(p,P,t)$, etc.  

\section{Gauge transformation not a canonical transformation}

In this section we shall prove that the gauge transformation (\ref{gauge_transf}) {\em is not} a canonical trasnformation. To do that, consider the momentum $P_i$
\begin{eqnarray}
P_i & = & \frac{\partial L'}{\partial \dot q_i} \nonumber \\
    & = & \frac{\partial}{\partial \dot q_i} \left ( L + \frac{d f(q,t)}{dt}\right) \nonumber  \\
    & = & p_i +  \frac{\partial}{\partial \dot q_i} \frac{d f(q,t)}{dt} 
\end{eqnarray}

Since, by definition $f=f(q,t)$, we have 
\begin{equation} \label{df}
\frac{d f(q,t)}{dt} = \sum_i \frac{\partial f(q,t)}{\partial q_i} \dot q_i + \frac{\partial f(q,t)}{\partial t}
\end{equation}
so that
\begin{equation}
\frac{\partial}{\partial \dot q_i} \frac{d f(q,t)}{dt} =  \frac{\partial f(q,t)}{\partial q_i}.
\end{equation}

Substituting this into (\ref{P_i}), we have
\begin{equation}
P_i =  p_i + \frac{\partial f(q,t)}{\partial q_i} 
\end{equation}

Now, from this we can write down the new Hamiltonian $H'$ as
\begin{eqnarray}
H'(q,P,t) & = & \sum_i P_i\, \dot q_i - L'(q,P,t) \nonumber \\
          & = & \sum_i \left (p_i + \frac{\partial f(q,t)}{\partial q_i} \right)\dot q_i-\left( L(q,p,t)+\frac{d f(q,t)}{dt}\right) \nonumber \\
	  & = & H(q,p,t) + \sum_i \frac{\partial f(q,t)}{\partial q_i}\dot q_i - \frac{d f(q,t)}{dt}
\end{eqnarray}

From (\ref{df}) we have that 
\begin{equation}
\sum_i \frac{\partial f(q,t)}{\partial q_i} \dot q_i - \frac{d f(q,t)}{dt} = - \frac{\partial f(q,t)}{\partial t}
\end{equation}	  
so that, finally
\begin{equation}
H'(q,P,t) = H(q,p,t) - \frac{\partial f(q,t)}{\partial t}
\end{equation}

The canonical equations of motion are
\begin{eqnarray}
\dot q_i & = & \frac{\partial H'(q,P,t)}{\partial P_i} \nonumber \\
         & = & \frac{\partial H(q,p,t)}{\partial p_i}\frac{\partial dp_i}{\partial dP_i} \nonumber \\
	 & = & \frac{\partial H(q,p,t)}{\partial p_i}
\end{eqnarray}
and
\begin{eqnarray}
\dot P_i & = & - \frac{\partial H'(q,P,t)}{\partial q_i} \nonumber \\
         & = & - \frac{\partial H(q,p,t)}{\partial q_i} + \frac{\partial }{\partial q_i}\frac{\partial f(q,t)}{\partial t}
\end{eqnarray}
which demonstrates clearly that the gauge transformation (\ref{gauge_transf}) is not a canonical transformation because it does not leave the Hamilton's equations of motion invariant under such a transformation, contrary to the conclusion reached by T.L.Chow in \cite{Chow}.

\section{Conclusion}

In this work we have demonstrated that, contrary to previously claimed assertion, the gauge transformation that leaves Euler-Lagrange equations of motion invariant, is not a canonical transformation since Hamilton's equations of motion are not preserved with such a gauge transformtation. Chow's mistake is in the definition of Hamilton's equation of motion where he took as 
\begin{eqnarray}
\dot q_i & = &  \frac{\partial H}{\partial \dot p_i} \nonumber \\
\dot p_i & = & - \frac{\partial H}{\partial \dot q_i} \nonumber 
\end{eqnarray}
while the correct expressions for Hamilton's canonical equations of motion are
\begin{eqnarray}
\dot q_i & = &  \frac{\partial H}{\partial p_i} \nonumber \\
\dot p_i & = & - \frac{\partial H}{\partial q_i} \nonumber 
\end{eqnarray}

\end{document}